\documentclass[twocolumn,aps,prb]{revtex4}
\usepackage{graphicx}
\usepackage{amsmath}
\usepackage{amsfonts}
\usepackage{amssymb}

\long\def\symbolfootnotemark[#1]{\begingroup\def\thefootnote{\fnsymbol{footnote}}\footnotemark[#1]\endgroup}
\long\def\symbolfootnotetext[#1]#2{\begingroup\def\thefootnote{\fnsymbol{footnote}}\footnotetext[#1]{#2}\endgroup}

\newcommand{\sub}[1]{\ensuremath{{}_\textrm{#1}}}
\renewcommand{\sup}[1]{\ensuremath{{}^\textrm{#1}}}
\newcommand{\tab}[1]{Table \ref{#1}}
\newcommand{\units}[1]{\ensuremath{\ \textrm{#1}}}

\newcommand{\nd}{\sup{nd}}

\newcommand{\st}{\sup{st}}

\begin{document}

\title{\bf Bandgap opening in metallic carbon nanotubes due to silicon impurities}

\author{Branden B. Kappes and Cristian V. Ciobanu\footnote{Corresponding author, email: cciobanu@mines.edu}}
\affiliation{Division of Engineering, Colorado School of Mines, Golden, Colorado 80401, USA}

\begin{abstract}
Controlling the bandgap of carbon nanostructures is key to the
development and mainstream applications of carbon-based nanoelectronic devices.
We report density functional theory calculations of the effect of silicon
impurities on the electronic properties of carbon nanotubes (CNTs). We have found
that Si adatoms can open up a bandgap in intrinsically metallic CNTs,
even when the linear density of Si atoms is low enough that they do not create an adatom chain along the tube.
The bandgap opened in metallic CNTs can range up to approx. 0.47 eV, depending on adsorption site, linear density of Si adatoms, and CNT chirality.
\end{abstract}

\maketitle

Much of the promise behind carbon nanotube-based electronic devices is the ability to control the electronic properties of systems with nanoscale dimensions. The high aspect ratio, strong covalent bonding, and long electron mean free path make CNTs ideal for ultra-thin interconnects; simultaneously, the ability to tune the electronic properties, such as the bandgap, through physical\cite{Louie2001, Yao2001, Poklonski2008} and chemical\cite{Rubio1996, Zhang2000, Liu2002} modifications provides a basis for creating remarkable small-scale devices. Although the idea of
bandgap engineering in CNTs is not new,\cite{Crespi1997, Louie2001, Yang2002, Miyake2005} realizing practical CNT-based devices is constrained by the lack of facile control over their electronic properties.

Chemical modification of CNTs via adsorption has a significant effect on their electronic properties, which can be understood through the impact that the adatoms have in the first Brillouin zone (BZ) of the CNT. This BZ is a set of line segments whose separation and number are related to the diameter, and whose length depends on spatial periodicity along the axis.\cite{Crespi1997, Saito1998} The BZ of a CNT is rotated with respect to that of graphene by an angle determined by CNT chirality. Ignoring tube curvature effects, the symmetry of the graphene and CNT reciprocal lattices permits the translation (folding) into the first BZ of graphene of the line segments of the nanotube BZ that extend beyond this zone. From this it follows that a CNT will only share the electronic properties of a graphene sheet that arise from the symmetry points that these two structures have in common.  Such is the origin of metallic, narrow- and broad-band gap semiconducting CNT; the BZ of the former passes through both the $K$ and the $K'$ of the graphene BZ, narrow-gap semiconductor CNT through the $K$ point only, and broad-gap CNT through neither.\cite{Crespi1997}  Breaking the symmetry of a CNT via adatoms provides a mechanism for tailoring the electronic properties of these graphitic materials.\cite{Zhang2000, Li2002, Liu2002, Zhang2009}

Investigations of the adsorption of various species has lead to the discovery of several new phenomena, including {\em semiconductor-to-metallic transitions} or functionality as spin polarizers for CNTs and graphene.\cite{Yang2002, Durgun2003, Lehtinen2003, Durgun2004, Durgun2006, Lehtinen2004, Chan2008, Mao2008} As the efforts for carbon-based nanoelectronics develop in a technological climate still strongly dominated by silicon devices, from a broader perspective it is useful to study the properties of nanostructures based on both carbon and silicon. Such studies can lead to novel applications based on tailoring the electronic properties of CNTs through doping with Si or through binding with Si surfaces and nanostructures. Here, we show that {\em metallic CNTs can become semiconducting} when covered with Si adatoms. Even when the atomic concentration of adsorbed silicon is as low as 2\% (i.e., lower than that necessary to form a bonded atomic chain along the nanotube), the doped CNT can develop a bandgap of up to 0.474 eV depending on the adsorption site, adatom density, and CNT chirality.

\begin{figure}[!htb]
  \begin{center}
    \includegraphics[width=7.2cm]{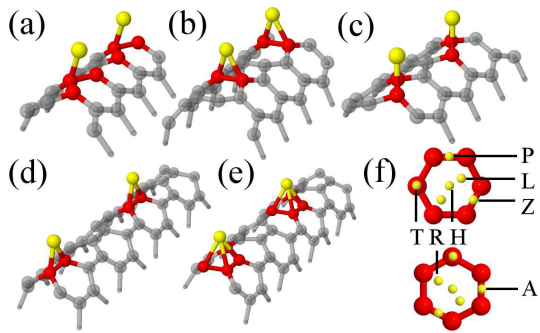}
  \end{center}
\caption{(Color online) Starting positions for the relaxation of a Si adatom on armchair (a-c) and zigzag (d,e) CNTs. Panel (f) shows the binding sites available on armchair (top) and zig-zag (bottom) CNT.}
  \label{fig:binding}
\end{figure}

The computational cells in our DFT calculations\cite{DFTcalcs} contain a CNT with $N$ unit cells along its axis and a single Si adatom. Figures \ref{fig:binding}(a-c) depict repeated supercells for (5,5) CNTs with $N=2$ and different initial Si adatom locations; similarly, Figs.~\ref{fig:binding}(d,e) show cutouts of supercells for (6,0) CNTs with $N=2$.  We set vacuum spacings of 12 \AA\ in
the directions perpendicular to the axis, and no vacuum spacing along the tube. Figure~\ref{fig:binding}(f) shows the binding sites available for the Si atoms on armchair and zigzag CNTs. The manner in which the electronic properties of CNTs arise from their chirality is well established;\cite{Hamada1992, Mintmire1992, Saito1992, Miyake2005} however, in an effort to isolate the properties that pertain to the transformation of a CNT from metallic to semiconductor, we examine four CNTs of {\em similar} diameters: two armchair [(4,4) and (5,5)] and two zigzag [(6,0) and (7,0)]. Both armchair CNTs are intrinsically metallic, while (7,0) is semiconducting. The (6,0) CNT is predicted via the zone folding approach to be a narrow gap semiconductor, but $\sigma^*-\pi^*$ rehybridization effects render this tube metallic.\cite{Li2001, Louie2001}

Depending on the chirality of the CNT, the Si adatoms relax preferentially over the P site (above the midpoint of a C-C bond) on armchair CNTs, and over the six-atom ring (sites H, R, or L) on zigzag CNTs [Fig.~\ref{fig:binding}(f)]. Our findings are summarized in \tab{tab:properties}, which includes the binding energies for the selected CNTs and adatom sites, as well as bandgap values. The binding energies we calculated are somewhat lower than those reported elsewhere for larger diameter CNTs, but the stable binding sites are consistent with previous reports.\cite{Durgun2003} The preferred binding site is thus insensitive to the diameter for the CNTs considered here and in Refs.~\onlinecite{Durgun2003}, ~\onlinecite{Lehtinen2003}, but not to the chirality.

Table \ref{tab:properties} further clarifies the relationship between the chirality of the CNT and the most stable binding site for Si. On the (4,4) CNT, the Si adatom breaks the highly strained C--C bond to form two, 1.9\units{\AA} long Si--C bonds. The formation of an adatom ``bridge'' is consistent with previous reports.\cite{Lehtinen2004} The binding energy of Si on the (5,5) CNT is substantially smaller despite the increase in tube diameter; this is likely due to the fact that the C--C bond is not broken in this case (thus no strain is released). In the absence of a transverse C--C bond, the Si adatoms on the (6,0) and (7,0) zigzag CNTs tend toward one of the three sites inside a six-atom ring (H, L, or R), and display a larger binding energy\cite{Durgun2003, Lehtinen2004} than the adatoms on the (5,5) armchair tube. It is interesting to note that the preferred adatom site depends on the density of Si atoms; with the exception of the (4,4) tube, the two most stable sites on the CNTs studied change with the distance between the adatoms (refer to Table~\ref{tab:properties}). This observation shows that even though the linear adatom densities are low ($<$2 Si atoms/nm) and no bonded adatom chain  is formed along the CNT, there are still significant strain-mediated interactions between the adatoms.

\begin{table}[!htb]
\caption{Stability and bandgap of selected CNTs with Si adatoms. The columns show the chiral vector $(n, m)$, the Si spacing $d_\textrm{Si-Si}$, the two most stable sites with their binding energies $E_b$, the bandgap and adatom position that leads to that gap.}  \label{tab:properties}
\begin{minipage}{0.5\textwidth}
  \begin{center}
  { % adjust table row/col separators
    \renewcommand{\arraystretch}{1.1}
    \renewcommand{\tabcolsep}{0.1cm}
    \begin{tabular}{c c c c c c | c c}
      \multicolumn{8}{c}{\rule{0pt}{2ex}} \\
      \hline
      \hline
      $(n,m)$ & $d$\sub{Si-Si} & 1\st & $E_b$\ & 2\nd & $E_b$ & Site, $N$ & Gap \\
            & (\AA)        &      & (eV)     &      &     (eV)     &          & (eV) \\
      \hline

      (4,4)
           & 4.95 & P                               &  2.100 & Z &  1.786 & R, 2 & 0.474 \\
      \multicolumn{6}{c|}{}                                            & Z, 2 & 0.259\\
           & 7.43 & P                               &  1.931 & Z &  1.694 \\
      \hline
      (5,5)
           & 4.95 & P                               &  1.219 & Z &  1.200  & R, 2 & 0.428 \\
      \multicolumn{6}{c|}{}                                            & Z, 2 & 0.235 \\
           & 7.43 & Z                               &  1.424 & R &  1.363 \\
      \hline
      (6,0)& 4.29 & R                               &  2.179 & Z &  1.785 & R, 1 & 0.290 \\
           & 8.58 & L                               &  2.377 & Z &  2.226 \\
           & 12.9 & Z                               &  2.249 & A &  1.978 \\
      \hline
      (7,0)
           & 8.58 & L                               &  2.120 & Z &  2.021  & A, 2 & 0.402 \\    %intrinsic & 0.233 \\

      \multicolumn{6}{c|}{}                                            & Z, 2 & 0.352 \\
      \multicolumn{6}{c|}{}                                            & L, 2 & 0.304 \\
           & 12.9 & Z                               & -1.998 & A & -1.584 & A, 3 & 0.356 \\
      \multicolumn{6}{c|}{}                                            & Z, 3 & 0.319 \\
      \hline
      \hline

      \multicolumn{8}{c}{\rule{0pt}{2ex}} \\
    \end{tabular}
    } % adjust table row/col separators
  \end{center}
  \end{minipage}
\end{table}

The adatom bonding and the ensuing tube-mediated interactions that occur for densities $\leq$ 2 adatoms/nm are responsible for severely altering the electronic properties of CNTs. Binding at certain sites is accompanied by opening bandgaps in intrinsically metallic CNTs, which can range up to 0.474 eV. Adsorption at site R induces a bandgap for all metallic tubes studied here (Table~\ref{tab:properties}) provided that the distance between adatoms remains smaller than 5\AA. This is particularly interesting for the (6,0) case, where the most stable adsorption site (R) is the only one that leads to a significant electronic gap (0.29 eV). The Si adatoms also affect the electronic structure of the semiconducting (7,0) tube, which in the absence of impurities has a gap of 0.233 eV; as seen in the Table~\ref{tab:properties}, adsorption at sites A, Z, or L increases the bandgap beyond 0.3 eV.

\begin{figure}
  \begin{center}
    \includegraphics[width=6.5cm]{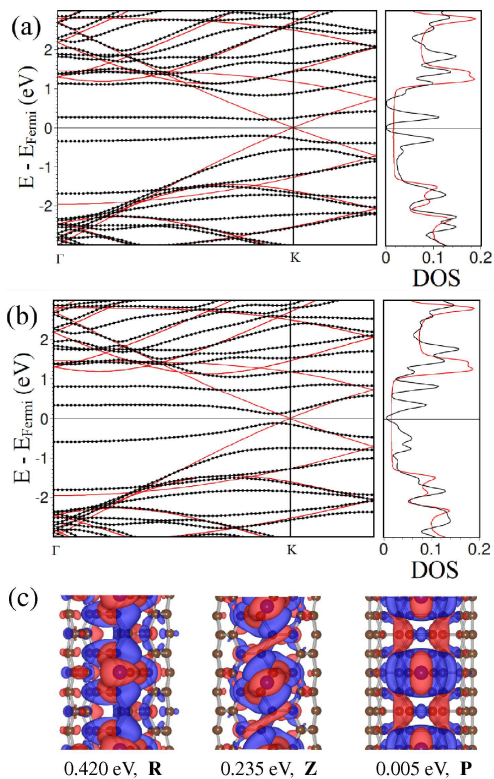}
  \end{center}
  \caption{(Color online) (a,b) Band structure (left) and density of states (DOS, right) for a (5,5) CNT with $N=2$ and the Si adatom placed at (a) site R and (b) site Z. The continuous red lines represent the band structure and DOS for the (5,5) tube with no adatoms. (c) Isosurfaces of the charge density transfer (red excess, blue deficit) for the (5,5) CNT with Si adatom at R, Z, and P sites, showing the correlation between the increased spatial symmetry of the electron transfer and the decrease of the bandgap.}
  \label{fig:xDB}
\end{figure}

To better illustrate the influence of Si adatoms on metallic CNTs, we have plotted in Fig.~\ref{fig:xDB}(a,b) the electronic structure and density of states (DOS) of Si-doped (5,5) CNTs for two adsorption sites (R and Z), in comparison to those of same CNT without adatoms. For both sites, adsorption of Si leads to significant band splitting and bending, as well as to heavy bands around the Fermi level which ultimately determine the bandgap of the doped CNTs. While the band structures in Figs.~\ref{fig:xDB}(a) and (b) show gaps of 0.428 eV and 0.235 eV, respectively, in the DOS plots these values appear much smaller due to the smearing employed to calculate the DOS.

For the same CNT and with the same linear density of Si adatoms (corresponding $N$=2), we have plotted the isosurfaces of the electron density transferred upon adsorption at sites R, Z, and P [Fig.~\ref{fig:xDB}(c)]. The transferred density was computed as the difference (taken at every point in space) between the density of the relaxed Si-doped CNT, and the densities generated separately by the CNT and the Si adatom. Depending on the adsorption site, the charge transfer can display different symmetries. When the symmetry group of the transferred density is $2\sigma_h$ (site R), the gap is 0.42 eV; as seen in Fig.~\ref{fig:xDB}(c), the value of the bandgap decreases with increasing symmetry of the charge transfer, reaching nearly 0.0 eV for site P. We have found this correlation between bandgap and transferred charge density for all the metallic nanotubes we investigated, and illustrated it in the Supplementary Information. The physical origin of this correlation lies in that the Si adatoms that generate low-symmetry charge transfers efficiently break the symmetry of the two carbon sublattices, which in turn leads to a bandgap. We have also verified this correlation for a more direct type of symmetry breaking, that in which the Si atom substitutes a C atom of the tube; in this case, for impurity spacings similar to those shown in Table~\ref{tab:properties}, the bandgap values are $\sim 0.15$ eV indicating that the adsorbed atoms can be more effective in opening a bandgap in metallic CNTs than Si substitutionals.

While we have shown that Si adatoms can induce metallic to semiconductor transitions in CNTs, the question remains of whether CNTs with active Si adsorbates are thermodynamically accessible. Similarities between diffusion behavior in graphene and CNTs suggest that procedures to control adatom concentrations on graphene by standard processes may also be applicable to CNTs.\cite{Lehtinen2003, Banhart1999} From \tab{tab:properties}, assuming a low-flux random deposition of Si adatoms, both the (6,0) and (5,5) CNT would realize a large number of Si adatoms at sites that can induce the metallic to semiconductor transition, the former populating over 90\% of these sites below 1500\units{K}--a soft upper limit for temperatures achieved on common heating stages.\cite{Banhart1999}

In conclusion, we have found evidence that silicon impurities in atomic concentrations as low as $\sim$ 2\% can induce a bandgap in intrinsically metallic CNTs, and can also modify the bangap of semiconducting CNTs. The value of the induced gap depends on the adatom density, adsorption site, and nanotube chirality. We have found that these correlations are not sufficient to fully determine the propensity of Si atoms to open a bandgap
in the electronic structure of metallic CNTs. In addition, the charge density transferred to/from the Si adatom upon relaxation plays a crucial role in breaking the symmetries of the two carbon sublattices: the higher the symmetry of the charge transfer (as qualitatively assessed from the symmetry groups of the spatial electron density transferred), the lower the value of the induced bandgap. Our results can be useful in the context of integrating CNTs with Si-based devices, particularly those in which silicon substrates would be used to tailor the properties of nanotubes arranged in specific epitaxial configurations.

{\em Acknowledgments.} We gratefully acknowledge funding from NSF under Grants No. CMMI-0846858 and No. CMMI-0825592, as well as computational support from the Golden Energy Computing Organization and from the National Center for Supercomputing Applications (Grant No. DMR-090121).

\newpage

\renewcommand{\thefigure}{S.\arabic{figure}}
\begin{figure*}[!htb]
  \begin{center}
    \includegraphics[width=18cm]{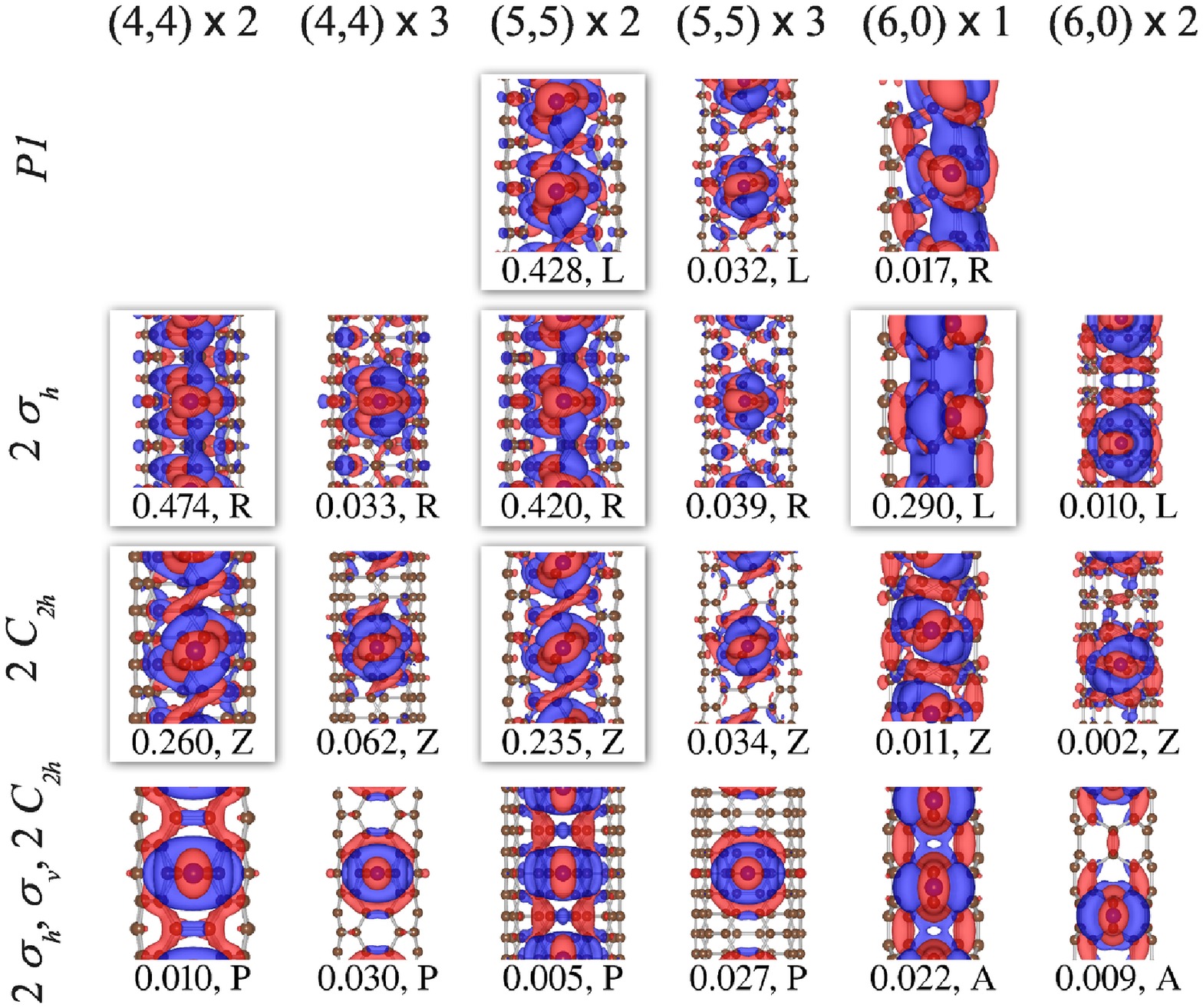}
  \end{center}
  \caption{{\bf Supplementary Information:} Relationship between the occurrence of electronic bandgaps (value in eV given below each panel, along with the adatom site) and the charge transfer upon Si doping of intrinsically metallic CNTs (isosurface plots at $\pm 0.001$ electrons/\AA$^3$). The CNTs are labeled $(n,m)\times N$, where $N$ is the number of primitive cells along the tube axis in the supercell. The formation of a significant bandgap in certain structures (framed, with drop shadow) depends on the CNT curvature, the binding site of the Si adatom, and the linear density of adatoms ($\propto 1/N$). However, these correlations are not sufficient to predict the formation of a bandgap. The spatial symmetry of the charge transfer --defined at each location {\bf r} as the difference in electronic densities of the doped system, the pure nanotube and the lone adatom $\Delta\rho ({\rm \bf r}) \equiv \rho_{\rm CNT + Si}({\rm \bf r}) - \rho_{\rm CNT}({\rm \bf r}) - \rho_{\rm Si}({\rm \bf r})$, provides additional insight into the trend of Si-doped metallic CNTs to form a bandgap: the higher the symmetry of $\Delta\rho$, the smaller the value of the bandgap. The high-symmetry regions of depleted (blue) and augmented (red) charge densities that exist when Si is located at the P or A sites inhibit the formation of a band gap, e.g., those systems that have all of the $2\sigma_h$, $\sigma_v$, and $2 C_{2h}$ symmetries. Conversely, atoms at the R and L sites (which lead to $2\sigma_h$ and $P1$ symmetries of the charge transfer) reveal a disparity in the charge density at the two carbon sublattices of the CNT and lead to the largest bandgaps.}  \label{fig:symmetry}
\end{figure*}

\end{document}